\documentclass[prl,twocolumn]{revtex4}

\usepackage{graphicx}
\usepackage{dcolumn}
\usepackage{bm}

\begin{document}

\preprint{ITP/UU-XXX}

\title{Sarma Phase in Trapped Unbalanced Fermi Gases}

\author{K. B. Gubbels}
\email{K.Gubbels@phys.uu.nl}
\author{M. W. J. Romans}
\author{H. T. C. Stoof}
\affiliation{
Institute for Theoretical Physics, Utrecht University,\\
Leuvenlaan 4, 3584 CE Utrecht, The Netherlands}


\begin{abstract}
We consider a trapped unbalanced Fermi gas at nonzero temperatures
where the superfluid Sarma phase is stable. We determine in
particular the phase boundaries between the superfluid, normal,
and phase-separated regions of the trapped unbalanced Fermi
mixture. We show that the physics of the Sarma phase is sufficient
to understand the recent observations of Zwierlein {\it et al}.\
[Science {\bf 311}, 492 (2006); Nature {\bf 442}, 54 (2006)] and
indicate how the apparent contradictions between this experiment
and the experiment of Partridge {\it et al}.\ [Science {\bf 311},
503 (2006)] may be resolved.
\end{abstract}


\maketitle

\emph{Introduction.} --- In the last two years impressive
experimental progress has been made in the field of ultracold
Fermi gases, in part due to the possibility to tune the
interatomic interaction strength by means of a Feshbach resonance.
This has led to the study of the crossover between the
Bose-Einstein condensation of diatomic molecules and the
Bose-Einstein condensation of atomic Cooper pairs, the so-called
BEC-BCS crossover \cite{Eagles,Tony,NSR}. Most recently, two
experimental groups have gone even a step further by obtaining
also full control over the polarization of the Fermi mixture. This
has allowed for the study of degenerate Fermi gases with
imbalanced spin populations, which is a topic of great interest in
many areas of physics ranging from condensed-matter physics to
nuclear and astroparticle physics. These pioneering experiments by
Zwierlein {\it et al}.\ \cite{Ketterle,Martin} and Partridge {\it
et al}.\ \cite{Hulet} induced a flurry of theoretical activity
\cite{Radzihovsky,Torma,Chevy,Pieri,Duan,
Erich,Masud,Ho,Demler,Machida,Torma2,Jani,Levin,Simons,Bulgac}.

The $^6$Li experiments of Zwierlein {\it et al}.\ and Partridge
{\it et al}.\ both revealed that superfluidity in an ultracold
Fermi gas is maintained upon going to an unequal mixture of two
spin states. However, rather contradictory results are obtained by
both experimental groups for the behavior of the Fermi mixture as
a function of the population imbalance. Zwierlein {\it et al}.\
observe a phase transition between the superfluid phase and the
normal phase at a high critical polarization of about 70\%,
whereas Partridge {\it et al}.\ seem to observe a transition
between two different superfluid phases at a low critical
polarization of about 10\%. One important aim of the present
Letter is, therefore, to propose a single theoretical picture in
which the qualitative differences in the observations by the two
experimental groups can be understood as two different sides of
the same coin. Based on this picture, we also make a detailed
quantitative comparison with the experiments of Zwierlein {\it et
al}.

\begin{figure}
\includegraphics[width=1.0\columnwidth]{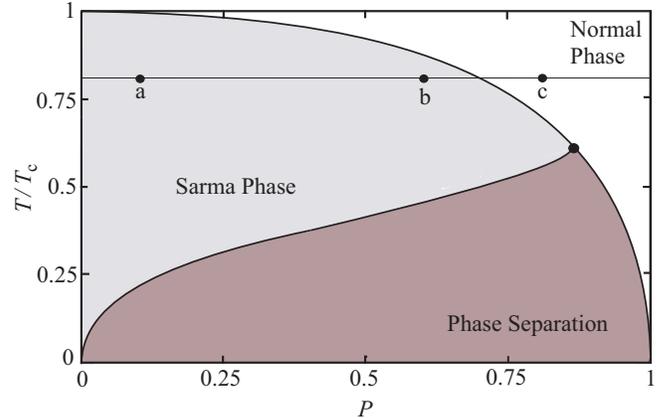}
\caption{\label{phasediagram} (Color online) Universal phase
diagram of a trapped unbalanced Fermi gas in the unitarity limit.
The polarization $P$ is given by $(N_{+}-N_{-})/(N_{+}+N_{-})$,
where $N_{\pm}$ designates the number of fermions in each
hyperfine state of the Fermi mixture. The temperature $T$ is
scaled with the critical temperature $T_c$ of the balanced Fermi
gas. The solid horizontal line gives the temperature that is used
for the comparison with the experiments of Zwierlein {\it et al}.\
presented in Figs.~2 and 3. The points a, b, and c correspond to
the polarizations used in the density profiles of Figs.~2a, 2b,
and 2c. \vspace{-0.25in}}
\end{figure}

The main results of our mean-field calculations are presented in
Fig.~1. Here we show the universal phase diagram of a trapped
Fermi gas in the unitarity limit as a function of temperature and
polarization. This phase diagram is universal in the sense that it
does not depend on the total number of fermions or the trap
geometry. Note that in determining the phase diagram we have
neglected fluctuations, which are known to be quantitatively
important in the unitarity limit. However, fluctuations are not
expected to alter the topology of the phase diagram in this case.
Fig.~1 reveals that there is a tricritical point in the trapped
Fermi mixture, which is well-known for the homogeneous case
\cite{Sarma,Mora,Simons}, but has to the best of our knowledge not
been studied yet for the harmonically trapped situation. In the
normal phase the gas is in its normal state throughout the trap.
In the Sarma phase the Fermi gas has a shell structure, in which
the core of the trapped gas is superfluid, whereas the outer
region is normal. Furthermore, the normal-to-superfluid transition
as a function of the position in the trap is of {\it second
order}. Since the superfluid order parameter $\Delta$ vanishes
continuously at the transition, we have for nonzero polarizations
always a region in the trap where $|\Delta|$ is so small that it
results in a gapless superfluid with negative quasiparticle
excitation energies, as first studied by Sarma \cite{Sarma}. Since
$|\Delta|$ increases towards the center of the trap, it is also
possible for small polarizations and low temperatures that the
superfluid becomes gapped in the center of the trap. This leads to
a gapped BCS superfluid core with a gapless Sarma superfluid and
normal shell surrounding it \cite{Shin}. Finally, in the
phase-separated region of the phase diagram, the superfluid core
and the normal shell of the gas are separated by a {\it
first-order} transition as a function of position, which implies
that $\Delta$ goes discontinuously to zero at a certain
equipotential surface in the trap.

Fig.~1 allows for a natural explanation of the qualitative
differences in the observations by the two experimental groups.
More precisely, we will argue in the second part of this Letter
that the experiments of Zwierlein {\it et al}.\ have observed the
transition from the normal phase to the Sarma phase, implying that
these experiments have been performed above the temperature of the
tricritical point. Moreover, we suggest that the experiments of
Partridge {\it et al}.\ have been performed in the temperature
regime below the tricritical point, since these experiments appear
to see the transition between a non-phase-separated and a
phase-separated superfluid phase. Note that this explanation is
fundamentally different from the proposal of Machida {\it et al}.,
who obtain a different phase diagram by considering the
Fulde-Ferrell-Larkin-Ovchinnikov (FFLO) phase, which has a
spatially varying superfluid order parameter, rather than phase
separation \cite{Machida}. As a result they obtain a Lifshitz
point instead of a tricritical point. They argue that both
experimental groups operate beneath this Lifshitz point, making it
difficult to explain the qualitative differences observed by both
groups. Moreover, there is presently no experimental indication
for the presence of the FFLO phase. Finally, we would like to
stress, that the local-density approximation that was used to
obtain the phase diagram in Fig.~1 is not sufficiently accurate to
describe all aspects of the experiments of Partridge {\it et al}.\
\cite{Erich,Masud,Demler}. For this reason we consider here from
now on only the experiment of Zwierlein {\it et al}., where a
quantitative comparison with our theory turns out to be possible.

\emph{Universal phase diagram.} --- To obtain the phase diagram of
Fig.~1 we use the mean-field theory for the Sarma phase in the
local-density approximation as described by Houbiers {\it et al}.\
\cite{Houbiers}. There the mean-field theory was applied to
superfluid ${}^6$Li in the BCS-limit. Here we incorporated the
relevant physics of the unitarity limit by using a generalization
of the approach put forward by Fregoso and Baym \cite{Baym}. As a
result, we end up with the following thermodynamic potential
\begin{eqnarray} \label{thermpot}
\Omega&=& \sum_{\mathbf{k}}\left( \epsilon_{\mathbf{k}} -\mu' -
\hbar \omega_{\mathbf{k}}+\frac{|\Delta|^2}{2
\epsilon_{\mathbf{k}}}
 \right) \nonumber \\
&-&\frac{1}{\beta}\sum_{\mathbf{k},\sigma}\ln(1+e^{-\beta
\hbar\omega_{\mathbf{k},\sigma}}) -\frac{1}{2}\sum_{\sigma}
N_{\sigma} \hbar\Sigma_{\sigma}~,
\end{eqnarray}
where the atomic dispersion is $\epsilon_{\mathbf{k}}=\hbar^2
\mathbf{k}^2/2m$ with $m$ the fermionic mass, $\Delta$ is the BCS
order parameter, $\sigma=\pm$ denotes the two hyperfine states of
the Fermi mixture, $N_\sigma$ is the number of atoms in each
hyperfine state, and $\beta=1/k_B T$. Moreover, the average
renormalized chemical potential $\mu'$ is given by
$\mu'=(\mu_{-}'+\mu_{+}')/2$ and the dispersions
$\hbar\omega_{\mathbf{k},\sigma}$ of the Bogoliubov quasiparticles
satisfy $\hbar\omega_{\mathbf{k},\sigma}=
\sigma(\mu_{-}'-\mu_{+}')/2 + \hbar\omega_{\mathbf{k}}$ with
$\hbar\omega_{\mathbf{k}}=
\sqrt{(\epsilon_{\mathbf{k}}-\mu')^2+|\Delta|^2}$. Finally, the
densities $n_{\sigma}=N_\sigma/V$ are given by
\begin{equation}
n_{\sigma} = \frac{1}{V}\sum_{\mathbf{k}} \left\{
|u_{\mathbf{k}}|^2 N(\hbar\omega_{\mathbf{k},\sigma})
+|v_{\mathbf{k}}|^2 [1-N (\hbar\omega_{\mathbf{k},-\sigma})]
\right\}~,
\end{equation}
where $V$ is the volume of the mixture and
$N(\hbar\omega_{\mathbf{k},\sigma})=1/[\exp(\beta\hbar\omega_{\mathbf{k},\sigma})+1]$
are the Fermi distributions for the Bogoliubov quasiparticles. The
BCS coherence factors $|u_{\mathbf{k}}|^2$ and
$|v_{\mathbf{k}}|^2$ are determined by the relations
$|u_{\mathbf{k}}|^2 = [1+ (\epsilon_{\mathbf{k}}-\mu')/\hbar
\omega_{\mathbf{k}}]/2$ and
$|u_{\mathbf{k}}|^2+|v_{\mathbf{k}}|^2=1$.

The renormalized chemical potentials are given by
$\mu_{\sigma}'=\mu_{\sigma}-\hbar\Sigma_{\sigma}$, where
$\mu_{\sigma}$ are the chemical potentials and the selfenergies
are approximated at unitarity by
\begin{equation}
\hbar\Sigma_{\sigma} = \frac{3\pi^2\hbar^2}{m} (\beta-\beta_{\rm
BCS})\sqrt{ \frac{\hbar^2}{m} \frac{1+\beta_{\rm BCS}}{2 \mu'}
}~n_{-\sigma}~.
\end{equation}
This expression can be understood as follows. Since the average
kinetic energy of the atoms involved in the interaction is given
by $2\mu'$, the effective interaction strength between the atoms
is expected to be proportional to $1/\sqrt{2\mu'}$ \cite{Baym}. To
understand also the proportionality constant we make use of the
fact that in the case of equal density and at zero temperature,
Eq.~(2) results in $\mu'_{\sigma}=(1+\beta_{\rm BCS})\epsilon_F$,
where we introduced the Fermi energy $\epsilon_F$ of a balanced
Fermi gas and we also recall that in the BCS theory $\beta_{\rm
BCS} \simeq -0.48$ \cite{Masud}. Substituting this result for
$\mu'_{\sigma}$ into Eq.~(3), we obtain for the chemical potential
$\mu_{\sigma}=\mu_{\sigma}'+\hbar\Sigma_{\sigma}
=(1+\beta)\epsilon_F$, which agrees with the exact result from
Monte-Carlo calculations for $\beta \simeq -0.59$
\cite{Georgini,Carlson}.

Taking the derivative of the thermodynamic potential with respect
to $\Delta^*$ and equating it to zero gives us the BCS gap
equation
\begin{equation} \label{gap}
\frac{1}{V}\sum_{\mathbf{k}}\left[
\frac{1-N(\hbar\omega_{\mathbf{k},+})-N(\hbar\omega_{\mathbf{k},-})}{
2\hbar\omega_{\mathbf{k}}} -\frac{1}{2\epsilon_{\mathbf{k}}}
\right] = 0~.
\end{equation}
Note that in deriving this gap equation we did not differentiate
the last term in the right-hand side of Eq. (\ref{thermpot}).
Differentiating also this term results in fluctuation corrections
to the mean-field theory on which we comment at the end of this
Letter.

\begin{figure*}
\includegraphics{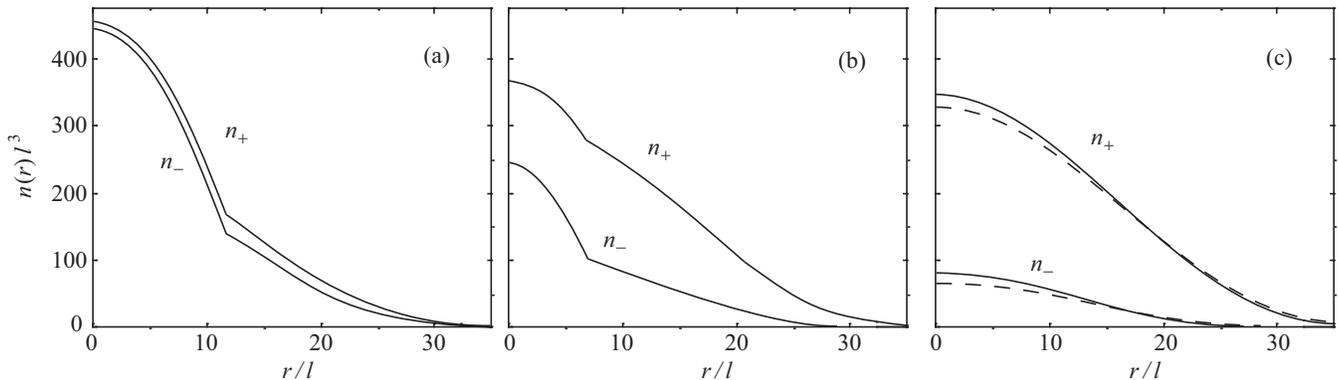}
\caption{\label{densprof} Typical density profiles for trapped
unbalanced Fermi gas above the tricritical point. The axial and
radial directions of the trap are scaled in such a manner that the
trap becomes effectively isotropic with a frequency
$\omega=(\omega_\perp^2\omega_z)^{1/3}$ and a size
$l=(\hbar/m\omega)^{1/2}$. The temperature and the polarizations
used for the three panels are indicated in Fig.~1. The total
number of atoms is $N = 1.4 \cdot 10^7$. The dashed lines in panel
(c) show the ideal gas results.\vspace{-0.1in}}
\end{figure*}

The above theory is valid for a homogenous Fermi mixture in the
unitarity limit. To account for the presence of an axially
symmetric trapping potential $V({\bf x}) = m (\omega_\perp
r^2+\omega_z z^2)/2$ we use the local-density approximation. This
means that the theory is locally homogeneous with a spatially
varying chemical potential. As a result, the local renormalized
chemical potential is given by $\mu_{\sigma}'({\bf
x})=\mu_{\sigma}-V({\bf x})-\hbar\Sigma_\sigma({\bf x})$. For a
balanced Fermi gas at zero temperature, we thus have
$\mu_{\sigma}-V({\bf x})=(1+\beta)\epsilon_F({\bf x})$ and
therefore we retrieve, as desired, the exact density profile in
this case. Note that in the outer region of the gas cloud the
renormalized chemical potentials become negative, so that $2
\mu'({\bf x})$ in Eq.~(3) is no longer a good measure for the
avarage kinetic energy of the interacting fermions. Therefore, we
then take $3 k_B T$ instead of $2 \mu'$ as an appropriate measure
for the kinetic energy.

We are now in the position to explain how we obtain the phase
diagram of Fig.~1. We first determine the line between the normal
and the two superfluid phases. This is achieved by solving the BCS
gap equation in the center of the trap and finding the temperature
at which the BCS order parameter vanishes. Inspection of the
thermodynamic potential reveals that the vanishing of the order
parameter can occur continuously or discontinuously, i.e., by a
second-order or a first-order phase transition. If the transition
is of first/second order, we go from the normal to the
phase-separated/Sarma phase. At the triciritical point these two
different kinds of transitions merge.

So far, we only looked at the center of the trap, but the
tricritical condition can also be satisfied at a certain
equipotential surface outside the center of the trap. This gives
us a point on the Sarma-to-phase-separation line. To see this,
consider a point on this line and raise the temperature slightly.
This changes the tricritical transition outside the center of the
trap into a second-order transition slightly closer to the center
of the trap, which means that the gas is in the Sarma phase. In a
similar way, a slightly lower temperature leads to a first-order
transition as a function of position in the trap, i.e., to the
phase-separated phase.

\begin{figure}[b]
\vspace{-0.1in}
\includegraphics[width=1.0\columnwidth]{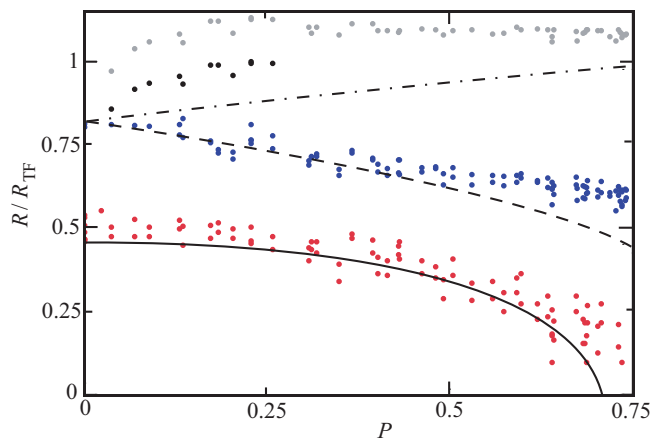}
\caption{\label{radii} (Color online) The radial size $R$ of the
Cooper-pair condensate (solid), and the minority (dashed) and
majority (dot-dashed) gas clouds as a function of polarization for
the temperature shown in Fig.~1. All radial sizes are scaled with
the radial Thomas-Fermi radius of the balanced Fermi gas $R_{\rm
TF}$. Also shown is the experimental data of Zwierlein {\it et
al}.\ \cite{Martin}. For the radial size of the majority cloud the
black points assume a hydrodynamic expansion, whereas the grey
points assume a ballistic expansion.}
\end{figure}

\emph{Comparison with experiment.} --- We now compare our theory
with the experiments of Zwierlein {\it et al}. First, we show in
Fig.~2 three typical density profiles of the gas: Two in the
superfluid Sarma phase and one in the normal phase. Using
different methods, similar density profiles have recently also
been obtained by Yi and Duan \cite{Duan} and Chien {\it et al.}
\cite{Levin}. The most striking feature in the density profiles is
the `bulge' in the minority and majority profiles in the Sarma
phase. This `bulge' is a direct consequence of the presence of the
condensate of Cooper-pairs and was indeed one of the most
important findings of the experiment. It shows that for an
unbalanced unitarity gas, in contrast to an unbalanced Fermi gas
in the BCS limit \cite{Houbiers}, the condensate of Cooper-pairs
has a very strong effect on the atomic density profiles. In the
normal phase the attractive effects of the selfenergies given in
Eq.~(3) also slightly increase the central densities of the two
species. However, this occurs always in a smooth featureless
manner whereas in the experiments of Zwierlein {\it et al.} a
distinct feature is seen in the majority density profile at the
edge of the minority cloud. This is maybe due to the fact that our
mean-field-like selfenergies are intended to exactly incorporate
the $\beta$ parameter of the balanced unitarity gas, but only
approximately account for the strong correlations in the normal
phase of the unbalanced unitarity gas. Another explanation could
be that in the experiment the density profiles are determined
after an expansion. This affects in particular the majority
density profile, since the outside of the majority cloud expands
ballistically, whereas the inner part inside the minority cloud
expands hydrodynamically and thus faster than the outer part. The
expansion is therefore not determined by a single scale factor and
this may lead to a pile up of atoms in the transition region.

Besides the above qualitative comparison, we can also make a
quantitative comparison with the experiment of Zwierlein {\it et
al}.\ by determining the radial size of the Cooper-pair condensate
and the radial sizes of the minority and majority gas clouds. The
radial size of the Cooper-pair condensate follows directly from
the point where $\Delta({\bf x})=0$, but the determination of the
minority and majority radii is somewhat more complicated because
we are working at a nonzero temperature and the density profiles
always have a gaussian tail. For simplicity we determine these
radial sizes by the conditions $\mu'_\sigma({\bf x})=0$, which
give the correct results at $T=0$, but underestimate the radial
size of the density profiles at nonzero temperatures. The results
of this procedure together with the experimental data are shown in
Fig.~3. In general, the agreement with experiment is very good,
which confirms our picture that the experiment is operating above
the tricritical point. As expected, at small polarizations the
radial size of the majority cloud agrees best with the
experimental data obtained by assuming a hydrodynamic expansion,
whereas for large polarizations our results approach the
experimental data obtained by assuming a ballistic expansion.

\emph{Discussion.} ---  The main remaining problem of our theory
is the high absolute value of the temperatures above the
tricritical point. The calculated temperatures are typically a
factor of 5 higher than what is found in the experiments. An
important consequence of the high temperature is that at the
Sarma-to-normal transition the densities of the two spin states
are not equal in the center of the trap in contrast to what is
found in experiments. The high absolute value of the temperature
is the result of neglecting fluctuations, that substantially shift
down the tricritical temperature. Including fluctuations in the
case of a balanced Fermi gas reduces the critical temperature in
the unitarity limit by a factor of 3, and fluctuations are
expected to be even more important in the unbalanced case. One
reason for the latter is that the Sarma phase is a gapless
superfluid, whereas the BCS phase at $P=0$ has a gap. However,
theoretically the study of fluctuation effects is rather
challenging for an unbalanced Fermi mixture at unitarity, since
the usual Nozi\`eres-Schmitt-Rink approach \cite{NSR} has some
unphysical features in this case \cite{Simons}. Therefore we are
developing a more advance theory including fluctuations. This will
determine the location of the tricritical point more accurately,
but, as mentioned previously, is not expected to alter the
topology of the phase diagram.

\emph{Acknowledgements.} --- We thank Randy Hulet and Martin
Zwierlein for stimulating discussions. This work is supported by
the Stichting voor Fundamenteel Onderzoek der Materie (FOM) and
the Nederlandse Organisatie voor Wetenschaplijk Onderzoek (NWO).

\end{document}